\documentclass[english]{article}
\usepackage{fullpage}
\usepackage[utf8]{inputenc}
\usepackage{color}
\usepackage{cite}
\usepackage{graphicx}
\usepackage{epsfig}
\usepackage{psfrag}
\usepackage{mathrsfs}
\usepackage{acronym}
\usepackage{amsfonts}
\usepackage{amsmath}
\usepackage{amssymb}
\usepackage{verbatim}
\usepackage{amsthm}
\usepackage{caption}
\usepackage{subfig}
\usepackage{url}
\usepackage{float}

\pdfoutput=1
\begin{document}

%\def\mb{\mathbf}
%\def\mc{\mathcal}
%\def\mbb{\mathbb}
%\def\e #1{\mathbb{E}\left\lbrace #1 \right\rbrace}
%\def\hl #1{\textcolor{red}{#1}}

%\title{LSA-Advanced and C-RAN: A (5G) Romance of Many Dimensions}
%\title{Could Census Tracts become a Disincentive for Sharing the 3.5 GHz band?}
%\title{Census Tracts - Disincentive for Sharing the 3.5GHz band?}
%\title{Are Census tracts license areas well-suited for Sharing the 3.5GHz band?}

% \author{\IEEEauthorblockN{Elma Avdic,~Irene Macaluso,~Hamed Ahmadi,~Ismael Gomez-Miguelez,~Laura Ingolotti,~Nicola Marchetti~and~Linda Doyle}
% \IEEEauthorblockA{CONNECT, Research Centre for Future Networks and Communications\\University of Dublin, Trinity College, Ireland\\
% Email: \{avdice,~macalusi,~marchetn,~linda.doyle\}@tcd.ie}
% }
%%\author{\IEEEauthorblockN{Elma Avdic\IEEEauthorrefmark{1}, Irene Macaluso\IEEEauthorrefmark{2}, Nicola Marchetti\IEEEauthorrefmark{3}, Linda Doyle\IEEEauthorrefmark{4}}
%% \IEEEauthorblockA{CONNECT, Research Centre for Future Networks and Communications\\University of Dublin, Trinity College, Ireland\\\IEEEauthorrefmark{1}avdice@tcd.ie, \IEEEauthorrefmark{2}macalusi@scss.tcd.ie, \IEEEauthorrefmark{3}marchetn@tcd.ie,  \IEEEauthorrefmark{4}linda.doyle@tcd.ie}}
%% <-this % stops a space
%
%%\maketitle

\title{LSA-Advanced and C-RAN: A (5G) Romance of Many Dimensions}

\author{Elma Avdic, Irene Macaluso, Hamed Ahmadi, Ismael Gomez-Miguelez, Laura Ingolotti,\\
 Nicola Marchetti, Linda Doyle\\ 
\\CONNECT, Research Centre for Future Networks and Communications,\\Trinity College Dublin, Ireland}
\date{December, 2014.}
\maketitle 

\begin{abstract}
We examine the dimensional limitations of Licensed Shared Access (LSA) regulatory framework for spectrum sharing and propose a solution for its more dynamic implementation. We consider an additional dimension for sharing, beyond those of time, space and frequency, i.e. the sharing of the infrastructure. We explain why we believe that the emerging regulations and use-case scenarios are limiting the potential of LSA, and with this in mind make a set of recommendations to unlock it fully. As a specific case study, we present our architectural work around Cloud-based Radio Access Network (C-RAN) and LSA to demonstrate the possibility of a dynamic implementation of multiple incumbents and multiple LSA licensees engaged in sharing. Our case study highlights multiple ways for LSA framework to evolve into a global and flexible sharing model, and prove that LSA and C-RAN are a good match to rethink how we look at networks targeting a more flexible, efficient and profitable way of operating them. 
\end{abstract}

%
%\begin{IEEEkeywords}
%%Spectrum sharing, geographic spectrum licensing, census tracts, 3.5 GHz band sharing framework, non-convex polygon optimisation, area loss, population per census tract with access to spectrum.
%\end{IEEEkeywords}

\section{Introduction}
%The pace of 5G technological change and innovation is accelerating and the surface of 5G capabilities is barely scratched. Intriguing problems and challenges of 5G are targeted to improve: spectrum usage efficiency, network capacity, cost efficiency and network resources utilisation.
%With the promise of enabling fibre-like speeds, high capacity demand for  Internet of Things (IoT) and ultra high-definition streaming video - it is clear that significant infrastructure upgrades will be required to enable the 5G forward vision.
%Challenges will need to be addressed by deploying both, evolutionary (e.g., Distributed Antenna Systems (DAS) or Massive Distributed MIMO (MD-MIMO)) and revolutionary (e.g., virtualisation, C-RAN) technologies to enable and support 5G. Modalities of spectrum sharing are so far explored for certain architectures (e.g., TV white space (TVWS)) or specific network technologies (e.g., LTE-A), but we argue that there is an active role for infrastructure in sharing to be considered as well.

It is possible to envision that time, frequency and space \textit{with} the infrastructure will act as key enablers of the spectrum sharing paradigm for 5G and beyond networks. These dimensions should \textit{intersect} and \textit{interact} to support more sophisticated forms of sharing, challenging the traditional ways we look at networks.

To unpack this statement, we compare the dimensional limitations of spectrum sharing with the \textit{Flatland} world of two dimensions by E. A. Abbott, where characters lack sight and senses to perceive the higher dimensions. In the world of spectrum sharing, the three-dimensional limitation imposed with time, frequency and space may be an obstacle to achieve more powerful solutions. We aim to explore the intersection of spectrum sharing dimensions by adding a further dimension into the pool - through sharing the infrastructure. One could argue that the infrastructure dimension adds to the complexity of an operator's network, but we believe that adding dimensionality into the problem increases the flexibility and helps to overcome limitations.

We put in the spotlight the infrastructure of a Cloud-based Radio Access Network (C-RAN) \cite{ChinaMob2011} coupled with a Distributed Antenna Systems (DAS) or Massive Distributed MIMO (MD-MIMO), where infrastructure elements are deployed in a coordinated way. We look beyond state-of-the-art of C-RAN and take a step further - to consider sharing and renting of the infrastructure resources to address the cost efficiency and resource utilisation challenges. Previous work \cite{IGM2013},\cite{IGM2014} points out that such a scenario offers network operators a higher degree of flexibility in pursuing cost efficiency trade-off in order to provide a service. Results showed that a virtual operator can transmit up to an order of magnitude more information at the same cost while relying on a trade-off between resources, as opposed to relying on using one resource only: more antennas or more additional Licensed Shared Access (LSA) spectrum from the pool to operate successfully. 
%C-RAN architecture gives us a flexible way to handle the pool of resources, based on spatial selection of  which resources to use in order to more effectively shape the network towards better performance.

%With today’s connected users evolving with their demands and expectations from service providers, the use of spectrum resources is evolving from having the exclusive use (licensed) or using the unlicensed (license-exempt) spectrum to sharing the exclusive license.
% At present, two or more systems/services can: 1) share the unlicensed spectrum in a TVWS framework with low level of QoS guarantees relying on the priority channels modality\cite{Singaporepaper}, 2) three-tier sharing \cite{PCAST}, \cite{GoogleSAS}, 3) share between licensed and unlicensed carriers via License Assisted Access (LAA)\cite{LAAconcept}, 3) share the exclusive license\cite{Rep205}.
 LSA, recently proposed as a sharing framework, introduces two tiers of users: incumbent (existing user in the band) and an LSA licensee (new entrant sharing the spectrum of an incumbent)\cite{Rep205}. The sharing arrangement between the tiers is based on a set of rules defined by the regulators aiming at the guaranteed level of QoS for both. In frequency bands where spectrum resources are under-utilised in time, space and/or frequency, the sharing happens in a binary mode - either incumbent uses the spectrum or a licensee.

%Opposed to its current interpretation, i.e. that of a sharing model, LSA is more a regulatory framework than a sharing technique. 
We aim to show how LSA can be implemented as a more general regulatory framework \cite{Rep205} for sharing. 
%So general that could open the door for \textit{multiple new entrants} engaging in the sharing agreement with \textit{multiple incumbents} while following the set of regulatory rules with a guaranteed level of QoS for both sides and through a certain licensing regime. 
In this paper, we are offering a way to move from one-to-one type of implementations to \textit{many-to-many} by bringing the infrastructure into the picture so that more dynamic scenarios of LSA (number of incumbents, number of licensees and the channels they operate on varying over time or space) can be implemented.

Our contribution is two-fold. The first step is to translate extracted information from a regulatory language into a more understandable form and to bring it together with the current LSA literature to explain how it lacks flexibility. 
%We make a set of suggestions and recommendations on how LSA sharing framework could be extended to enable more flexible and more dynamic implementations. 
We suggest directions for the modified regulatory scheme of LSA: LSA-Advanced framework. The second step is a case study with the architecture that will allow us to showcase the active sharing of infrastructure and spectrum in a multiple incumbents and multiple LSA licensees interaction.
%We propose a direction towards a global sharing model encompassing more flexible allocations, incentive systems, licensing regimes and diversity of 5G users and sharers.
%Spectrum should be available for reuse and re-licensing more often than today by rethinking the act of classification within a hierarchical, tiered model. Therefore,  

We build our story around the concept of \textit{extending the dimensionality to overcome limitations}. In this paper, we look beyond LSA and beyond C-RAN current state-of-the-art to bring together these two concepts in a highly dynamic, heterogeneous, virtualised network accommodating a diverse range of players, users, resources and technologies to shape the future networks for an expressive, competitive market.
\section{LSA Today}
\label{sec:SoA}
%The most well-known form of spectrum sharing that exists today is in the ISM bands, which operate on a \textit{license-exempt} basis. 
LSA regulatory framework promotes the idea of sharing on licensed basis rather than using a free-for-all commons approach and is complementary to both extremes of having and not having the license to spectrum rights.
% Following up the industrial initiative the  has published its first report, with the goal of extending the ASA model to a more generic one. 
In response to the Radio Spectrum Policy Group (RSPG) report \cite{RSPGreport}, several high level regulatory deliverables, defining the LSA sharing framework and its standardization, have been published by the ECC, EC and ETSI \cite{Rep205},\cite{ETSIlsa}.

The regulators see the LSA concept as a two-step framework of (1) frequency allocations and (2) frequency assignments/authorisations \cite{ETSIFM52}. During step one, the allocation is applied to a specific incumbent without prejudging the modalities of step two, namely what systems and services will be authorised as sharers (new entrants). According to the regulatory details, the LSA sharing framework is also supposed to be agnostic to the choice of access technology, network deployment or application of spectrum usage. This general approach opens the way for new entrants and different kind of actors e.g., virtual operators as service providers who will compete in the market.
\\ The LSA license is envisaged and defined as an individual right of use, delivered by the Administration to the LSA licensee. 
%Sharing under the LSA umbrella is binary by its nature: either incumbent or LSA licensee utilise the spectrum, having the exclusive spectrum rights of use at the specific time and/or location, with predictable QoS.
%For incumbents willing to enter the sharing agreement, it is important to be aware of potential sources of benefits as well as the downsides. 
LSA can be seen as a means of an incumbent holding the long-term primary rights of use of a band that it might otherwise need to vacate, were it not willing to share. Obviously these benefits need to be balanced by the risk of the spectrum resources being unavailable for periods of time in specific geographical areas. Thorough examination of regulatory deliverables suggests that there is the possibility of compensation from the users availing of the LSA scheme\cite{Rep205} opening space for a more comprehensive incentive system. 
%However, it seems to be considered that spectrum trading of any kind should not be allowed as a way for an incumbent to be incentivised properly.
%\\
%\textcolor{red}{add here incumbent protection and evacuation time text and change the paragraphs below!!}

%The LSA licensee has the benefit of getting additional spectrum through the LSA approach with a guaranteed QoS.

Currently, the LSA architecture reference model is under development \cite{RRSworkshop}. It is based around the concept of an LSA Repository and an LSA Controller, communicating through a reliable path and standardised interface. The LSA Repository will support the entry and storage of the sharing arrangement information and incumbent’s protection requirements. The LSA Controller manages access to the spectrum.

Getting access to spectrum resources through  LSA requires a mindset change for the licensee(s), in that they must rely on  not exclusively owned spectrum, and as a result the licensees are dependent on incumbent usage statistics. Sharing may also mean increased operational and technical complexity. In principle, the LSA licensee can hold on to the incumbent’s spectrum as long as the incumbent does not use it. In reality, incumbent’s usage in the band can be unpredictable, depending on the nature of its application. 
The potential need to evacuate the spectrum in case the primary holder needs it back is where in essence, LSA and exclusively licensed spectrum differ. An evacuation request is sent by the incumbent, via the LSA Database to the LSA Controller and passed to the OA\&M in traditional RAN deployments. OA\&M then configures and coordinates the network, vacating the spectrum according to the request (graceful exit of the band or faster response time required in urgent situations \cite{ETSIReq}). Sharing options between the incumbent use (in 2.3GHz) and  Wireless BroadBand (WBB) use are described in \cite{Rep56}, but technical conditions and implementations of sharing under LSA should be described at national level, considering the diverse incumbent use in 2.3GHz across CEPT countries.

% and is recommended to be located in the licensee’s network. This would give a licensee the right to protect the location of its own resources. 
%There will be an interface between Repository and Controller.

In principle the LSA framework is generic enough to allow: 1) different kinds of service providers as licensees and 2) for the LSA frequency choice not to be linked to a specific band. In practice however, EU trials \cite{LSAtrial} and regulatory deliverables \cite{Rep205}, \cite{ETSIlsa}, \cite{ETSIReq}, \cite{Rep56} currently interpret and implement the  LSA framework for: 1) traditional MNO acting as LSA licensee and 2) 2.3 GHz band. 
%\hl{This is a first step of LSA research in Europe with the emerging trials \cite{LSAtrial} and setting the regulatory framework for sharing \cite{Rep205} to shape the standardisation activites on LSA.} 
Following closely LSA research in the EU and engaging actively in discussions with the EU regulation and standardisation bodies, points to the idea that exploring more dynamic, flexible, open and more general LSA scenario could be the next step.

% Also, frequency choice for LSA so far is linked to a specific band through the use case trial in 2.3 GHz\cite{LSAtrial} and regulatory deliverables  band in the EU limited to a specific band, as it is currently the case for 2.3 GHz band in EU  Much of research in Europe and the emerging trials assume this to be the case, but it is only a first step towards considerations of LSA for many different IMT bands, identified for potential harmonisation.

%Following closely the regulatory and standardisations activities since LSA emerged and engaging actively in discussions with EU regulations points to the idea that exploring more dynamic, more flexible, more open and more general LSA scenario could be a next step after first LSA implementations are done. 

With this in mind, we frame our suggestions and recommendations in Section~\ref{sec:LSA-A} section as features of the LSA-Advanced sharing framework. 
%The vertical type of sharing arrangement between a non-MNO incumbent and MNO licensee should not be the only option for sharing under the LSA umbrella. There should be possibilities for more dynamic scenarios, with multiple players engaged in sharing. There should be users holding all types of licenses under this umbrella with a proper incentive system developed for all of them.

\section{Re-envisaging LSA for 5G and beyond}
\label{sec:LSA-A}
For those who fundamentally believe in spectrum sharing, LSA is a move in the right direction as it challenges the notion that incumbents who do not make optimal use of their frequency allocation can simply sit in these frequency bands without consequence. It is a move towards a \textit{share it or lose it} perspective and therefore much welcome from a sharing perspective. However, we contend that the emerging regulations and use-case scenarios are limiting the potential of LSA and with this in mind make the following set of recommendations.
\\

\begin{itemize}

\item LSA should not only be interpreted and implemented as a one-to-one relationship between the incumbent and sharer but extended to enable \textit{many-to-many} type of relationship. We see signs of this move in the recent work of ETSI RRS group\cite{RRSworkshop}, considering a one-to-many type of incumbent-licensee interaction. Its dynamic interpretation could be set for multiple incumbents and multiple licensees engaging in the sharing arrangement, with flexible rules, adaptable to the spectrum environment in a cognitive radio (CR) fashion to get through the benefits of sharing.

\item In the 5G context, it is crucial the LSA licensee is not simply seen as a traditional MNO (as in most of the current use-case scenarios) but rather that \textit{other types of systems and services can share}. There is a huge potential in the LSA framework to become a market oriented model in a wider setting than at present - by using its principle of sharing the exclusivity as an advantage. 

\item Exclusive assignments, given for a long time, which are also renewable are the main cause of spectrum underutilisation. The LSA framework currently recommends long-term licenses to provide security to the MNOs \cite{Rep205}, but LSA should support highly dynamic as well as more static longer term sharing scenarios. 
%The future systems will have to be able to operate under different authorisation modes (and across various bands) to unlock access to as much spectrum as possible.  

 \textit{Flexible licensing regime} (a mixture of long-, medium- and short-term models) would put different types of licensees under the LSA umbrella – those who seek multi-year assurance, those who operate on a shorter time scale (up to 3 years or less) and those who rent the rights of use for few minutes during rush hour or couple of hours of an important event. How successful such solutions would be depends on transaction costs, regulatory framework and limitations of sharing rules. 
\item LSA is currently a two-tier system, consisting of an incumbent and a licensee with specific well-defined access rights. We contend that the three-tier approach based along the lines of the PCAST framework for the 3.5GHz band in the U.S. is much more powerful \cite{PCAST}, \cite{GoogleSAS}. It allows for a third level of users, akin to the ISM band license exempt approach with additional interference constraint to protect the first two tiers. The three-tier approach is more comprehensive and allows for future expansion and can of course scale down to a two-tier approach depending on how conditions are set. The \textit{LSA-Advanced framework} comes with the backward compatibility drawing on the idea explained above. 
%It does not change the architecture or the interfaces between LSA system blocks, as depicted in Fig.~\ref{fig:anynetwork}.

\item Different \textit{mechanisms for assignment} should be considered ranging from simple rule based assignments to auctions held at specific intervals to dynamic spot-auctions. The number of actors who can play a role in competitive auctions is limited when long-term, expensive licenses are issued. There should be a secondary market within LSA scheme for which there is no space in its current regulatory framework. An example of a shorter term and cheaper license purchaser is a small scale operator, not in need for a big coverage area.

\item The potential of a \textit{hybrid licensing model} could be explored to modify the LSA scheme. Administrative licensing (command and control approach, e.g., public safety, military)), rights of use model (e.g., users with economic perspective in mind) and a license-exempt model (e.g., low power, short range devices) could come together in the LSA framework through the diversity of users and sharers scattered around the tiered model. 
%Licenses for users whose spectrum demand can be satisfied on a first come, first served basis (relays, satellite earth stations) could be LSA-mixed with the users who use their spectrum rationally, with economic perspective in mind (operators). Low power, short range devices (unlicensed) could enter the LSA pool as well, to use the public commons whenever higher priority tiers are not using it. 
In this dynamic setting, incumbent’s license could be awarded as a license based on actual use.

\end{itemize}
There is much work to be done on both the technical and policy side to make the above recommendations a reality. Here, we begin by tackling the coordination of multiple incumbents and multiple licensees to better make use of shared resources. We see this as a fundamental starting point and a way in which the wider vision for a future LSA system can be driven forward.

\section{Unlocking the Potential of LSA - A C-RAN based case solution}

%\begin{figure}
%\centering
%	\includegraphics[width=5in]{LSAandoperators_resized}
%	\caption{LSA interface with \textit{any network} operator}
%	\label{fig:anynetwork}
%\end{figure}

The pace of 5G technological change and innovation is accelerating and the surface of 5G capabilities is barely scratched. 
%Intriguing problems and challenges of 5G are targeted to improve: spectrum usage efficiency, network capacity, cost efficiency and network resources utilisation.
With the promise of enabling fibre-like speeds, high capacity demand for  Internet of Things (IoT), connected cars and ultra high-definition streaming video - it is clear that significant infrastructure upgrades will be required to enable the 5G forward vision.
%5G is supposed to be looked at from the level of wide range of \textit{use cases} it will encompass, rather than looking from the level of specific \textit{application} or \textit{technology} deployed or \textit{capability} or a \textit{requirement}. 
Performance of the \textit{network of networks} will be driven by user experience, leveraged by wide range of identified use cases.

Our case study is focused around high density, high and varied usage demand, low mobility users. This type of scenario (we refer to it as \textit{cloud-surfing}) might map well to large sporting events (e.g., Olympics or the Superbowl) or to a specific location such as future hospital supporting many connected health services. Virtual operators will make use of whatever infrastructure and spectrum they need to provide the varied services to the subset of users. We use the term \textit{supercell} to describe the area of coverage, a term associated with MD-MIMO.

% This is the type of scenario that might map well to large sporting events (e.g the Olympics or the Superbowl) or to a specific location such as a future hospital supporting many connected health services. 
% We see the varied services being provided by different virtual operators. These operators will make use of whatever infrastucture and spectrum they need to provide the service to the subset of users they are addressing. We use the term supercell to describe the area of coverage, a term associated with MD-MIMO.}
%Many antennas of C-RAN operator are distributed in the scenario setting location, PMSE incumbents would be broadcasting the event, i.e., olympics. 
%In this dynamic scenario (we call it \textit{cloud-surfing}), virtual operators would provide different services to many users scattered around a supercell \cite{IGM2014} and PMSE applications of portable cordless cameras or wireless cameras mounted on a helicopter would be using their spectrum as well. 

\subsection{C-RAN Today}

C-RAN is a recently proposed network architecture characterised by (a) the centralization of the basestation (BS) processing functions in a common location, (b) the use of distributed remote radio heads (RRH) linked by high-speed links to the processing centre and (c) the use of Cloud computing principles for the pooling of the computing resources. This centralization makes the C-RAN suitable for supporting cooperative techniques such as joint scheduling, or beamforming. In addition, its elasticity and scalability allows to dynamically shape the network (e.g. number of used antennas, processing resources, etc.) as a function of economical or environmental parameters.
C-RAN as currently envisaged and operated in various jurisdictions is the property of a single MNO. %The C-RAN operator may allow virtual mobile network operators (MVNO) to use the infrastructure in the same way MVNO uses existing MNO infrastructure today.

\subsection{Moving to More Extreme Sharing}

%5G will deal with different capabilities and technologies in order to cope with the range of identified use cases. 
%The capability of virtualization in 5G can help mobile carriers serve new devices and demands in 5G ecosystem, improving resource utilisation and management giving more flexible way to handle the pool of resources, from the perspective of application and service. By a operator's network perspective, the market driven approach and current trends in telecom world are pointing to the fact that more and more operators are becoming virtual, examples of even MNOs (e.g., AT\&T) not owning all of their infrastructure (towers) are quite common.

\begin{figure}
\centering
	\includegraphics[width=4in]{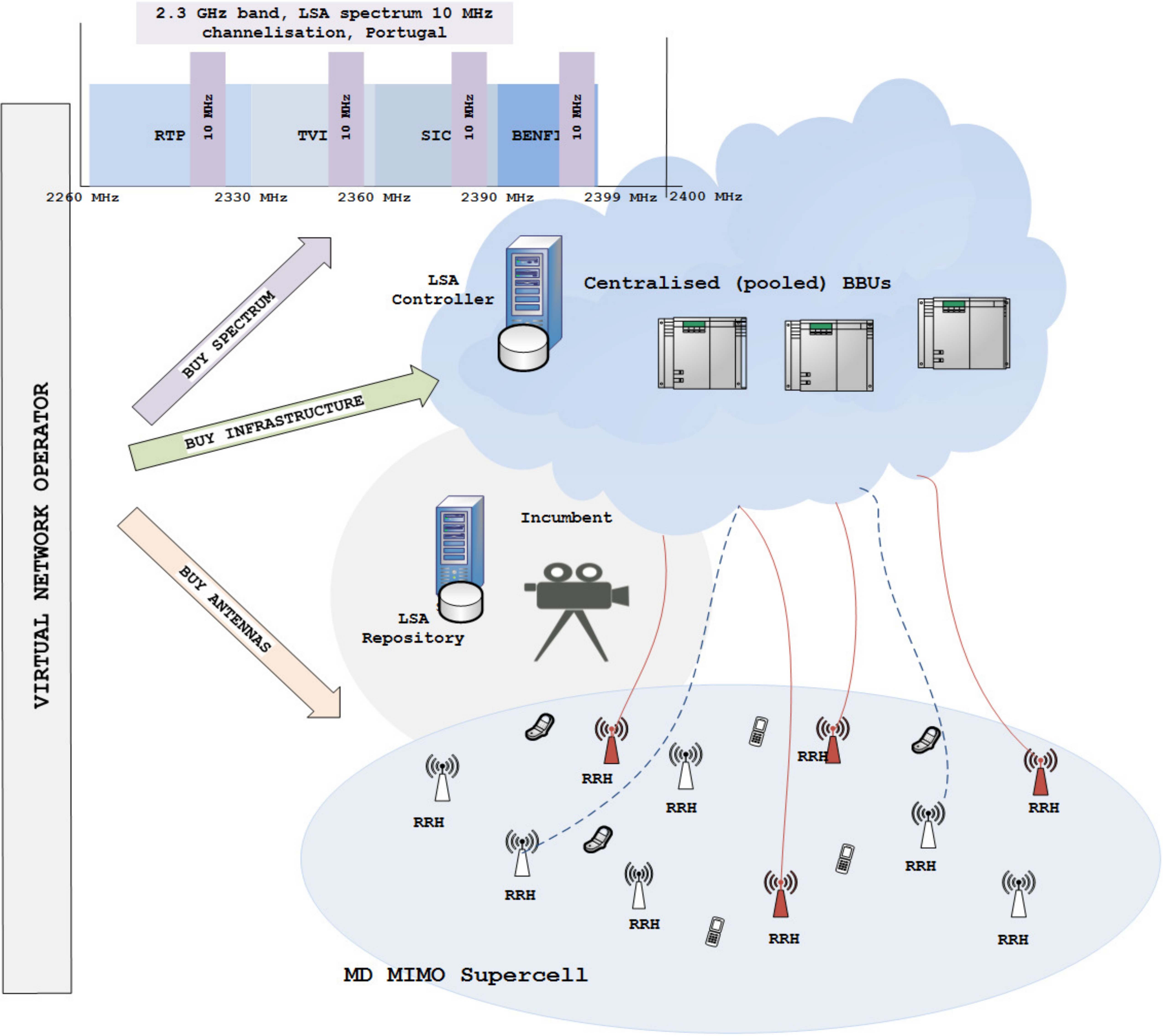}
	\caption{C-RAN network deploying LSA sharing framework}
	\label{fig:scenario}
\end{figure}

A C-RAN based case solution combines futuristic views of LSA and C-RAN. It tackles the 5G challenges of: \textit{improving spectrum efficiency} (via LSA-Advanced and MD-MIMO), \textit{resource utilisation} (flexible architecture), \textit{cost and energy efficiency} (C-RAN), \textit{5G capabilities} (virtualisation), \textit{technologies to improve network capacity} (C-RAN with MD-MIMO), \textit{use case} (hotspot coverage of a stadium), \textit{the pooling of resources, demands and services} (service-driven networks concept). 
%\begin{figure}
%\centering
%	\includegraphics[width=4in]{CRAN_LSA_withLSAspectrum_new-eps-converted-to}
%	\caption{C-RAN network deploying LSA sharing framework}
%	\label{fig:scenario}
%\end{figure}
\\We see networks evolving \cite{NWOB} to become dynamic entities composed out of pool of consumable commodities, many of which will be shared and where the  boundary between pure
software and hardware resources is blurred. This pool of commodities consists
of highly heterogeneous resources: information, infrastructure, knowledge, content, people, goods, RRHs, spectrum, backhaul, hypervisors, storage, processing power, etc. 
%\\A future C-RAN architecture combined with an LSA system for accessing spectrum goes to the heart of highlighting what we mean by this.
\\Fig.~\ref{fig:scenario} captures the ideas and concepts involved: distributed RRHs, owned by different entities (typically, C-RAN is owned by a single entity, i.e. the MNO), are scattered around a geographical area. The architecture includes a common pool of process and storage resources. Our aim is to show that this ‘pool perspective’ that takes a more extreme view of sharing can make significantly better use of the resources. In Fig.~\ref{fig:scenario} therefore, the virtual operators (LSA licensees) will rent or share the infrastructure according to their service and users demands (IoT, Over the Top (OTT), mobile broadband, voice\&data). In essence, they rely on a trade-off between renting more antennas or more spectrum to generate revenue.

\begin{figure}
\centering
	\includegraphics[width=6in]{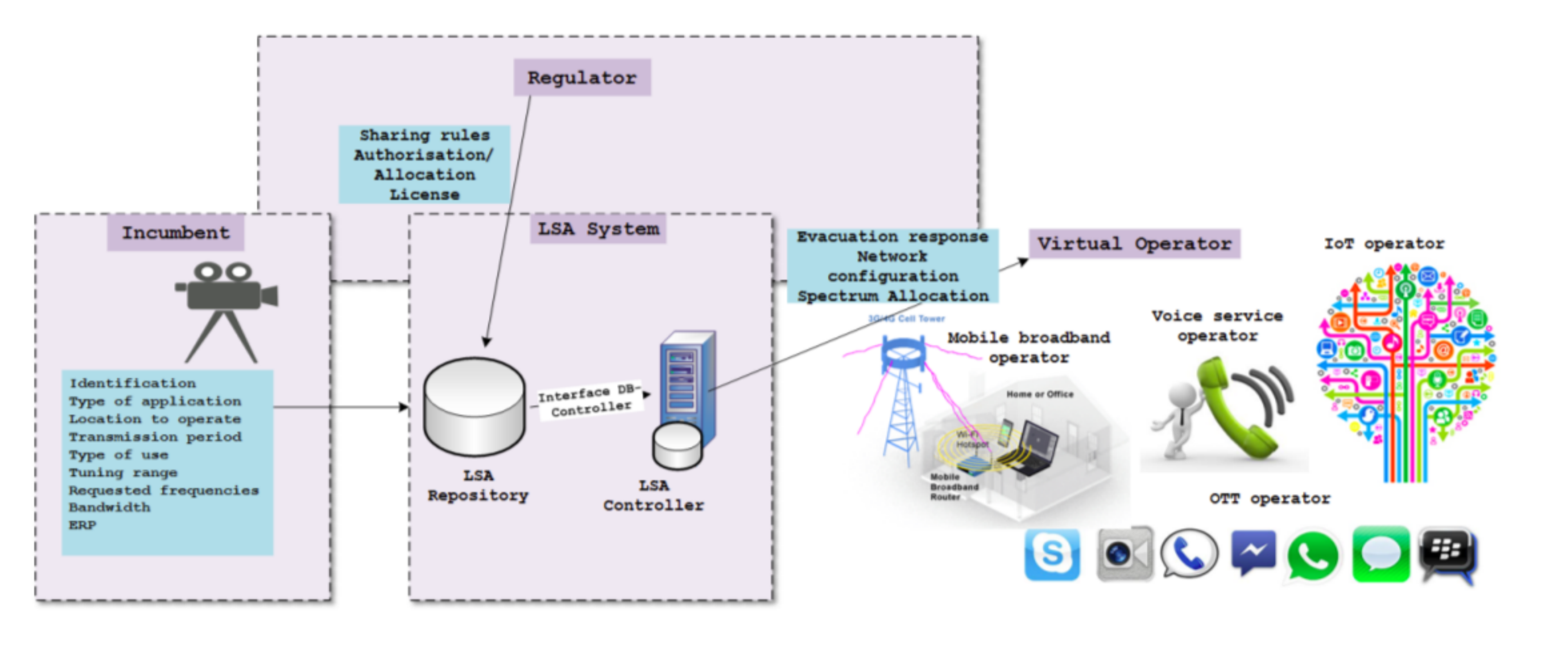}
	\caption{LSA interface with \textit{any network} operator}
	\label{fig:anynetwork}
\end{figure}

Added to the mix is a pool of spectrum made available through an LSA licensing scheme
and is provided by the PMSE incumbent (common incumbent usage in 2.3 GHz band across CEPT countries). The LSA Repository stores the information on incumbent spectrum availability, sharing arrangement between the virtual operators and incumbents, and regulatory rules they follow to operate in coexistence. The LSA Controller is located in the Cloud, performing the resource allocation and updating real-time network information which virtual operators may have access to. In the specific use case, we considered four incumbent TV stations, whose spectrum allocations in 2.3 GHz band and channelization is depicted in the Fig.~\ref{fig:scenario}. 
%\\Virtual operators access the resources dynamically, when they need them. 
\\Network resources are accessed dynamically, according to the need. Other resources are left for other players to use \footnote[1]{According to the ITU definition, virtual mobile network operators do not own the spectrum license nor infrastructure. However, here we refer to virtual operator or a service provider, as any entity that does not fully own all the resources necessary (spectrum, infrastructure, computational etc.) to deliver a given service.}. The concept does not affect the LSA architecture or the interfaces between LSA system block. They remain the same, as depicted in Fig.~\ref{fig:anynetwork}. 
%The architecture we are exploring gives virtual operators (IoT, Over the Top (OTT), mobile broadband, voice\&data) the possibility to access resources dynamically, when they need them. Other resources of the network are left for other players to use \footnote[1]{According to the ITU definition, virtual mobile network operators do not own the spectrum license nor infrastructure. However, here we refer to virtual operator or a service provider, as any entity that does not fully own all the resources necessary (spectrum, infrastructure, computational etc.) to deliver a given service.}. The concept does not affect the LSA architecture or the interfaces between LSA system block. They remain the same, as depicted in Fig.~\ref{fig:anynetwork}. 
%\\ In a service-driven networks approach, the resources that any type of operator (IoT, Over the Top (OTT), mobile broadband or voice\&data operator) needs to perform a service are: spectrum resources coming from the LSA pool of resources (the LSA license holder gets the access to them, i.e. virtual operator), infrastructure resource coming from infrastructure provider (in this case it is C-RAN operator, but can be any operator), computational resources coming from the cloud etc. It is a highly dynamic, heterogeneous, virtualised network accommodating a more diverse range of players.

Prior to analysing the benefits of the proposed architecture in Section~\ref{sec:study}, we briefly discuss its implications for the coexistence issues in the LSA framework: the dynamic evacuation of the LSA band when the incumbent claims the spectrum back and the protection of the incumbent from interference.

In the case of traditional RAN deployments, where baseband processing is  integrated in a base station, responding to an evacuation request  in a particular region causes a certain delay, as shown with the world's first LSA trial\cite{LSAtrial}. OA\&M communicates with each eNB active in the region where the incumbent needs spectrum back \cite{Rep56}. For the C-RAN deployment, the response to an evacuation request is done in one control point - in the Cloud. In fact, the C-RAN
operator reacts to the evacuation request by stopping transmission of IQ data from baseband unit (BBU) pool to RRHs via high speed, low latency fibre link. In case the C-RAN operator and the LSA licensee are not the same entity, as in our case study, the only additional messaging in the  evacuation information flow is between the virtual operator (LSA licensee) and the C-RAN operator (notifying C-RAN about the evacuation request received from incumbent), drawing on the idea that the LSA license holder is responsible for the
interference.

%The fundamental issues of LSA framework applied in a coexistence of two or more different systems/services engaged in sharing the spectrum are: protecting the incumbent from the interference and dynamic evacuation of the LSA band when incumbent claims the spectrum back. 
%\\ Responding to evacuation request (either graceful shutdown or offloading the traffic to another carrier in traditional RAN deployments) in C-RAN deployment is done in one control point – in the Cloud. In traditional RAN deployments radio and baseband processing are integrated in a base station. Shutting down the active base stations in a particular region to stop the transmissions and return the spectrum to the incumbent causes a certain delay in the network (as shown with world's first LSA trial \cite{LSAtrial}, done via OA\&M communicating with each eNB active in the region where incumbent needs spectrum back \cite{Rep56}. Based on the contract on service agreement made beforehand between C-RAN operator and virtual operator, C-RAN operator reacts to the evacuation request by stopping transmission of IQ data from BBU pool to RRHs via high speed, low latency fibre link. Additional messaging in this evacuation information flow with C-RAN network is only between virtual operator and C-RAN operator (notifying C-RAN about the evacuation request received from incumbent), drawing on the idea that the LSA license holder is responsible for the interference. 

 System requirements for LSA describe the technical conditions for sharing and coexistence \cite{Rep56} in terms of adjacent and co-channel interference thresholds. Antennas are deployed in our remote network in MD-MIMO fashion. With the spatial diversity of channel conditions and a narrow beams feature, the interference that may affect incumbent sharing in a particular region with the licensee, is reduced. The control of interference is much easier because of centralised BBU pool taking into account the contribution of each RRH \cite{CRANoverview}.

% This addresses the coexistence issue of different incumbent and licensee systems sharing in a particular region. 
%As for interference, we should note that the control of it is much easier because of centralised BBU pool taking into account the contribution of each RRH and the narrow beams facilitated by the MD MIMO.

% Radiated power is divided by many antennas, which almost annihilates the adjacent channel interference. 
%In terms of inter-cell interference, antennas in this scenario form a supercell so the cell loses its meaning as in conventional systems and hence, large amount of control signalling needed to avoid inter-cell interference is eliminated. Consequently, the latency due to this type of control signalling is not present. 
%\\ Perhaps it is also worth noting that C-RAN with massive MIMO can be seen as lower overall potential interference source (to coexist with another system through sharing), compared to traditional networks - since it enables the sophisticated interference mitigation techniques such as Coordinated MultiPoint (CoMP) and Enhanced Inter-cell Interference Coordination (eICIC) to be facilitated, due to the concept of pooling BBU resources \cite{CRANoverview}. 

%In general, control of the interference is much easier because of centralised BBU pool taking into account the contribution of each RRH and the narrow beams facilitated by MD MIMO.tential of LSA - A C-RAN based case solution}
\section{Case study} \label{sec:study}

We considered  mobile virtual network operators (MVNOs) that rent antennas, fibre and processing resources from the infrastructure provider, C-RAN operator. The MVNOs pay for the resources in a pay-per-use fashion and they share the infrastructure. Indeed, while the LSA spectrum cannot be shared between multiple MVNOs, it is possible for multiple MVNOs transmitting in different spectrum bands to share an RRH, provided that the RRH is equipped with a multi-channel or wide-band transceiver. Processing resources cannot be shared, but we assume they are unlimited. The underlying assumption is that MVNOs and incumbents interact in a many–to-many basis. This way, when one or more incumbents return, the available resources can be reallocated to meet the MVNOs requirements.

We simulated a scenario consisting of $5$ MVNOs requesting resources from the C-RAN operator to serve their users with a minimum rate requirement. We consider a centralized resource allocation scenario, where resources are allocated so as to maximize the revenue, defined as the
difference between the income and the operational costs. The operational costs are a linear function of the number of antennas and used bandwidth, and the income is proportional to the rate provided to every MVNO. Resources are allocated to an MVNO only if its rate requirement is satisfied. In line with the channel restriction rules defined in the LSA framework, the resource allocation scenario only considers multiples of $5$ MHz channel as possible spectrum assignments. The proposed approach exploits the concurrent allocation of diverse and partially interchangeable resources, namely spectrum and antennas, to maximize the revenue. To quantify the capabilities of a dynamic C-RAN infrastructure in enhancing the sharing features of LSA we compare the performance of the proposed approach with the performance obtained when the pool of shared
resources only contains the LSA spectrum and each MVNO uses a static C-RAN, i.e. a C-RAN with a fixed number of antennas.

\begin{figure*}
\centering
\subfloat[Case I, minimum rate 200 Mbps ]{\includegraphics[width=6.5cm]{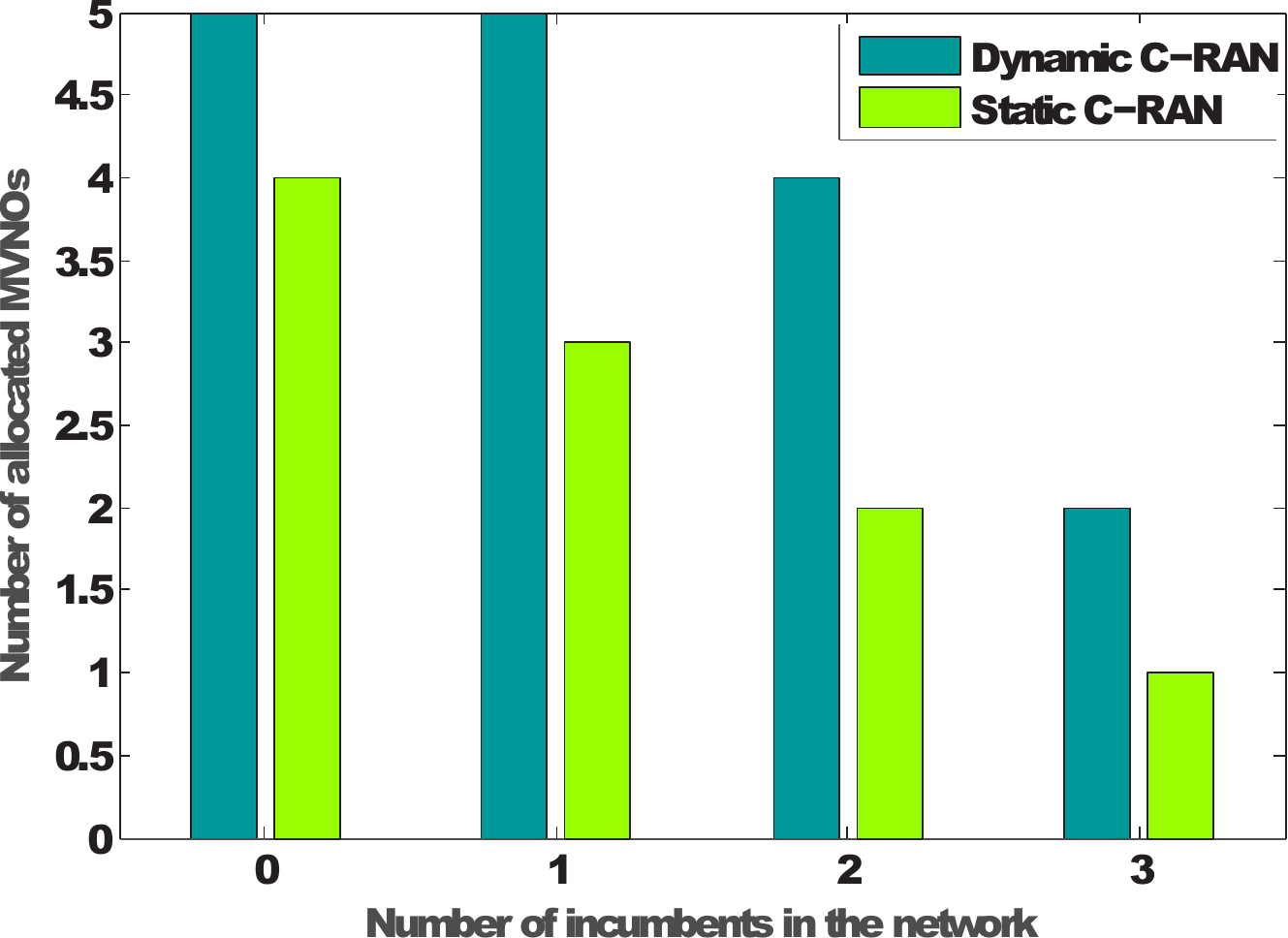}
\label{results1}}
\hfil
\subfloat[Case II, minimum rate 430 Mbps]{\includegraphics[width=6.5cm]{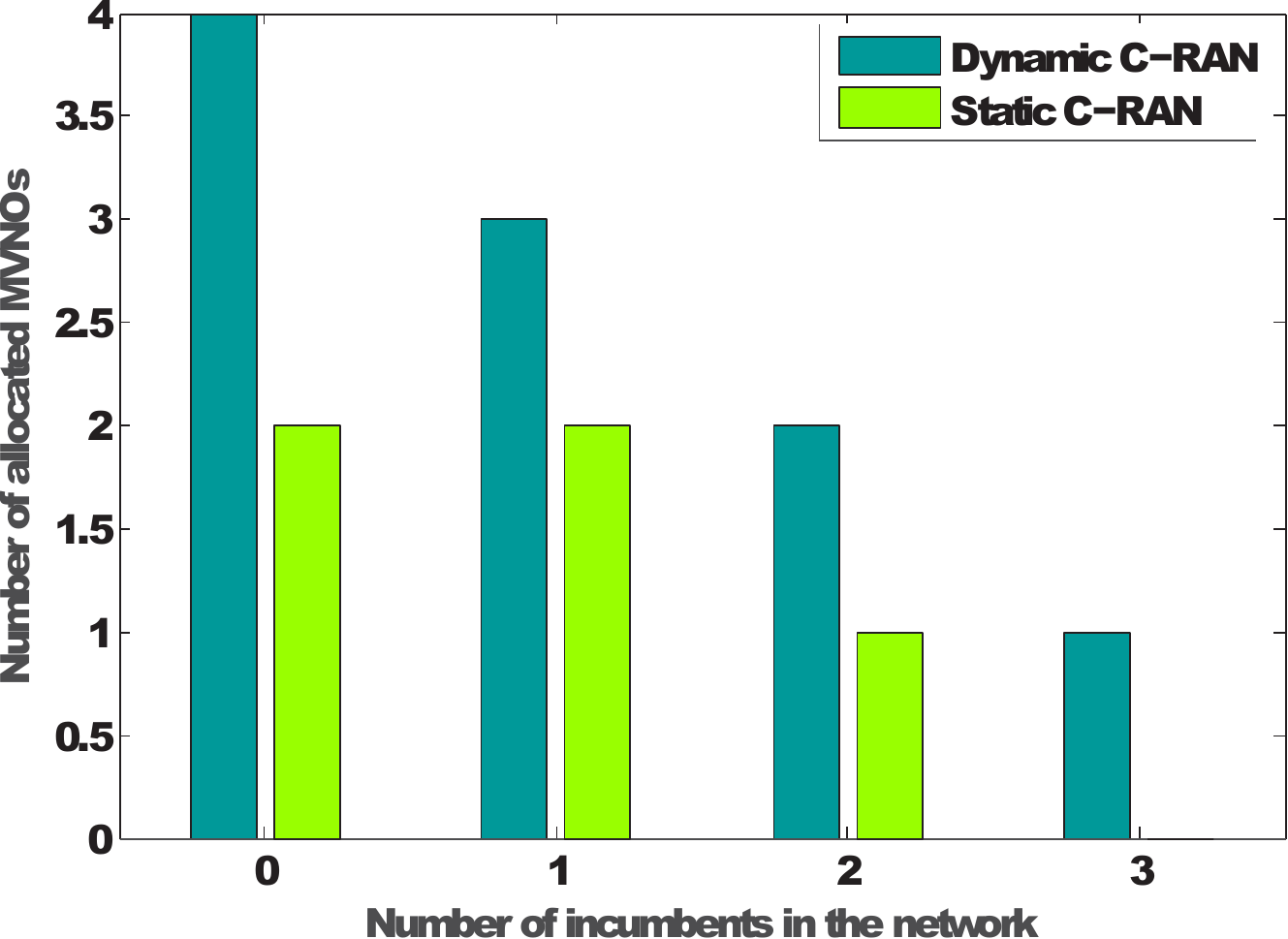}
\label{results2}}
\caption{Served virtual operators, function of number of incumbents in the network}
\label{fig:results}
\end{figure*}

We considered the total available bandwidth to be 40 MHz when no incumbent is in the network, and it decreases by $10$ MHz for each incumbent that comes back. The minimum number of antennas used by the dynamic C-RAN per MVNO is $20$ and it is the same as the number of antennas used by the static C-RAN system. The maximum number of antennas in the dynamic C-RAN system is $100$. It is worth noting that the overall number of antennas in the two systems is the same and the difference in performance is due the combined flexibility provided by sharing infrastructure and spectrum and the fact that average rate per user increases with the number of antennas.

In Fig.~\ref{fig:results} we compare the maximum number of MVNOs that can be served by the two systems as the incumbents return. In this case we considered a very high price per rate, so that the revenue is maximized when the maximum number of MVNOs is served, according to the available resources and minimum required rate of each problem instance. The results show that the dynamic C-RAN system can make better use of the available resources especially when the required minimum rate for each MVNO is higher. It should also be noted that, for a lower minimum rate requirement (see Figure 3.(a)), the dynamic C-RAN system is able to accommodate all $5$ MVNOs even in the presence of $1$ incumbent.
By exploiting the additional flexibility  of the dynamic C-RAN  infrastructure, each MVNO can meet the minimum rate requirement using only $5$ MHz. This means that up to $8$ MVNOs could be allocated when no incumbent is present. For the static C-RAN case, each MVNO requires a $10$ MHz channel to satisfy the same rate requirement. Hence, in the presence of $1$ incumbent, only $3$ MVNOs can be allocated. Fig.~\ref{fig:results} also shows that the maximum number of served MVNOs decreases as minimum rate increases.

\begin{figure*}
\centering
\includegraphics[width=5in]{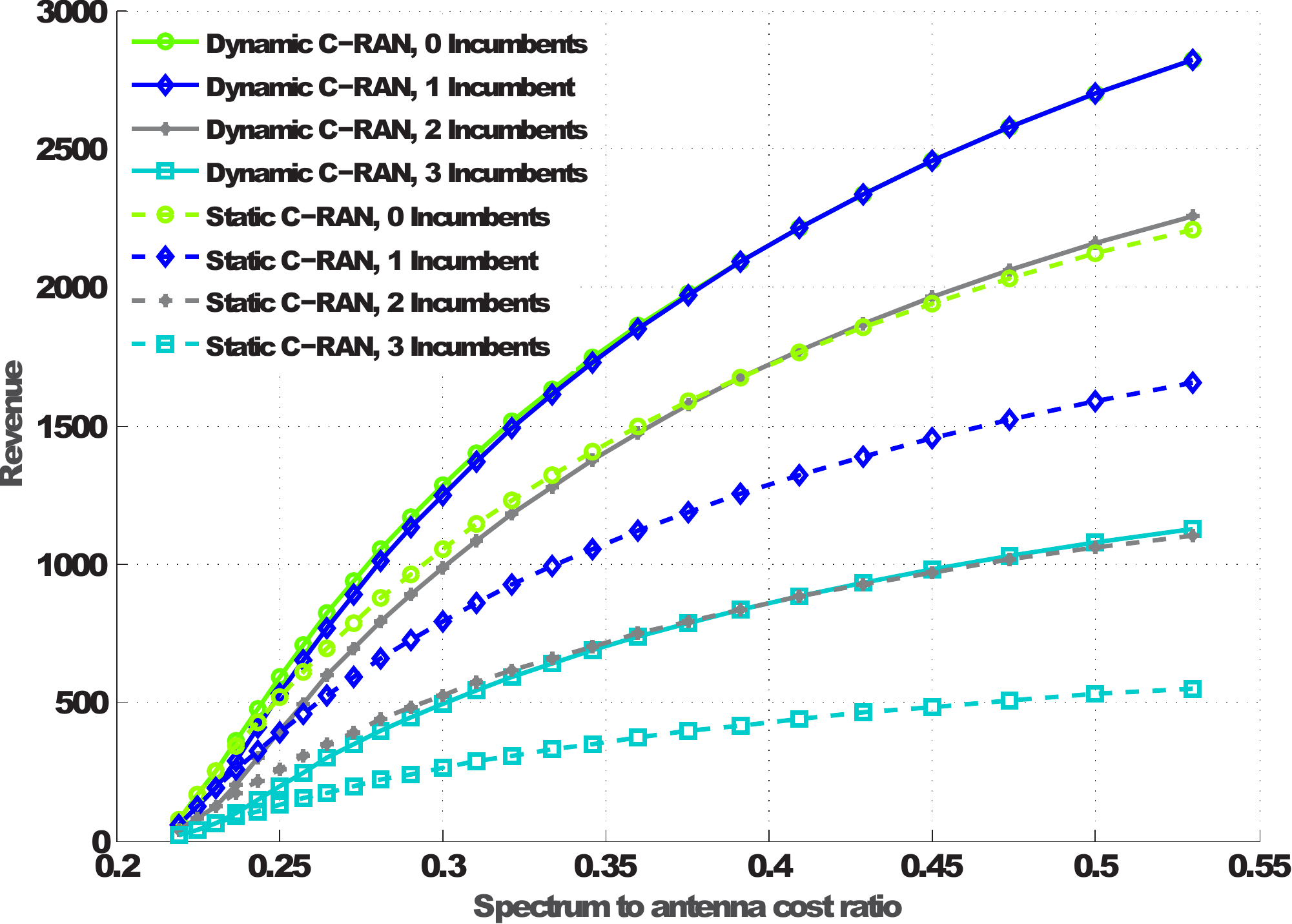}
\caption{Revenue as a function of cost when minimum rate is 200 Mbps}
\label{fig:revenue}	
\end{figure*}

In Fig.~\ref{fig:revenue} we investigate the effect of the spectrum to antenna cost ratio on the revenue of the two C-RAN systems. Results in Fig.~\ref{fig:revenue} refer to a fixed spectrum cost. Our results show that as the ratio increases, the dynamic C-RAN system outperforms the static C-RAN system significantly. When the cost of antennas increases, i.e. when the ratio decreases, the difference in revenue of the two systems decreases to zero. This happens when the dynamic C-RAN system uses the minimum possible number of antennas due to their higher cost. This in turns results in a lower number of allocated MVNOs. It is important to point out that the trend observed in Fig.~\ref{fig:revenue} only depends on the spectrum to antenna cost ratio and not on the absolute values of spectrum and antenna costs. The actual spectrum cost value only determines the point where the revenues of both systems are equivalent.

\section*{Evolution towards a global, flexible model}
\label{sect:Conclusions}
The LSA way to share the spectrum seems to be controllable and reliable, but is it flexible? 
 LSA rules in EU regulation are driven by a situation in a particular band (i.e. 2.3 GHz) and LSA is being interpreted as a way to unlock additional spectrum for MNOs. It does not allow different licensing regimes and rights of use under its umbrella (other than individual rights of use based on exclusivity). It does not design incentive system for tiers and does not allow unlicensed users to engage in sharing. 
We laid out a set of regulatory and technical suggestions to overcome these limitations.
 
   As a specific case study, we analysed a highly flexible sharing model that combines a futuristic view of LSA with a futuristic view of C-RAN. In particular, our study relies on a many-to-many type of incumbent-licensee interaction. This licensing scheme in combination with the additional flexibility provided by the C-RAN shared infrastructure allows to exploit the available resources more efficiently. 
   Hybrid licensing model would go beyond long-term licenses for new entrants (the main justification for operators to invest and engage in sharing). The shared infrastructure concept with C-RAN MD-MIMO reduces the investment costs and therefore enables the possibility of shorter term licenses. 
   
The model we analysed is based on a two-tier system, as it is currently envisioned by the LSA framework. As noted in Section \ref{sec:LSA-A}, we think that the LSA framework should evolve to a three-tier model, following the example of Spectrum Access System (SAS) - based sharing of 3.5GHz band, \cite{PCAST}, \cite{GoogleSAS}.
With the LSA-Advanced model we aim to address the features from both models (e.g., rethinking the act of classification within a hierarchical, tiered model; analysing the 'access rights' of unlicensed users, considering the value of shared spectrum when tiers interact in different combinations and extending this case study on architectural work further). Next steps are set towards identifying the technical implications of the differences between two models and comparison of their current implementations in the U.S. and EU. This will open the way for policy implications analysis and potentially overcome most of the limitations we discussed.
%We will focus on detailed examination of interaction between three tiers, scaling down the model according to the benefits and downsides for incumbents and licensees engaged in sharing. 
From the regulation and implementation perspective, LSA-Advanced superset would provide a way for using spectrum more efficiently in order to \textit{dynamically respond to the needs of the service}, with the demand on more of everything. It would also be an opportunity for European and U.S. regulatory and standardisation bodies to combine the two- and three-tier implementation experiences into a comprehensive regulatory and technical spectrum sharing model, applicable globally.

\section*{Acknowledgement}
This work is supported by the Seventh Framework Programme for Research of the European Commission under grant ADEL-619647 and the Science Foundation Ireland under grant CONNECT 13/RC/2077. We would like to thank Dr. Tim Forde for his insightful comments and valuable feedback which improved the quality of this paper.
%
%\bibliographystyle{IEEEtran}
%\bibliography{IEEEabrv,CTrefs}

\begin{thebibliography}{15}

%\bibitem{abbott1884}
% Edwin A. Abbott,
%  \emph{Flatland: A Romance of many Dimensions},
%  Seeley and Co., London,
%   1884.
   
 \bibitem{IGM2013}
Ismael Gomez-Miguelez, 
\emph{Radio and computing resource management in SDR clouds}, 
Ph.D. dissertation, Universitat Politecnica de Catalunya, Barcelona,
Spain, 2013.

\bibitem{IGM2014}
Gomez-Miguelez, I., Avdic. E. Marchetti, N., Macaluso, I., Doyle, L., 
\emph{Cloud-RAN platform for LSA in 5G networks - Tradeoff within the Infrastructure},
6th International Symposium on Communications, Control and Signal Processing (ISCCSP), 2014.

\bibitem{ChinaMob2011}
China Mobile Research Institute,
\emph{C-RAN: The Road Towards Green RAN}, 
white paper, 2011.

\bibitem{RSPGreport}
RSPG,
\emph{Report on Collective Use of Spectrum (CUS) and other spectrum sharing approaches},
European Commission, Radio Spectrum Policy Group, 
2011.

\bibitem{Rep205}
ECC Report 205,
\emph{Licensed Shared Access (LSA)}, 
2014.

\bibitem{PCAST}
President’s Council of Advisors on Science and Technology (PCAST) Report,
\emph{Realizing the Full Potential of Government-Held Spectrum to Spur Economic Growth},
2012.

%\bibitem{Singaporepaper}
%Ser Wah Oh, Chin Choy Chai,
%\emph{Geo-location database with support of quality of service for TV White Space},
%24th International Symposium on Personal Indoor and Mobile Radio Communications (PIMRC),
%2013.

%\bibitem{LAAconcept}
%3rd Generation Partnership Project, 3GPP,
%\emph{Technical Specification Group Radio Access Network; Study on Licensed-Assisted Access to Unlicensed Spectrum;(Release 13)},
%2014.

%\bibitem{}
%Matinmikko, M.; Okkonen, H.; Palola, M.; Yrjola, S.; Ahokangas, P.; Mustonen, M., "Spectrum
%sharing using licensed shared access: the concept and its workflow for LTE-advanced networks,"
%Wireless Communications, IEEE, vol.21, no.2, pp.72-79, April 2014.

\bibitem{LSAtrial}
M. Palola, M. Matinmikko at al.,
%Hartikainen, L. Tudose, A. Kivinen, J. Paavola, and K. Heiska,
\emph{Live field trial of Licensed Shared Access (LSA) concept using LTE network in 2.3 GHz band},
in Proc. IEEE DySPAN, McLean, VA, USA,
2014.

\bibitem{GoogleSAS}
Google Access Services,
\emph{Spectrum Access System: Managing Three Tiers of Users in the 3550-3700 GHz Band}, 
available at \url{http://wireless.fcc.gov/workshops/sas\_01-14-2014/panel-1/Marshall-Google.pdf}.

\bibitem{RRSworkshop}
ETSI RRS group,
\emph{ETSI workshop on Reconfigurable Radio Systems - Status and novel Standards} available at
\url{http://www.etsi.org/news-events/events/807-etsi-rrs-workshop-2014}.

\bibitem{ETSIlsa}
ETSI TR 103 113 V1.1.1 (2013-07),
\emph{Electromagnetic compatibility and Radio spectrum Matters (ERM; System Reference document (SRdoc); Mobile broadband services in the 2 300 MHz - 2 400 MHz frequency band under Licensed Shared Access regime},
2013.

\bibitem{ETSIReq}
ETSI TR 03 154 V1.1.1, 
\emph{Reconfigurable Radio Systems (RRS);System requirements for operation of Mobile Broadband Systems in the 2 300 MHz- 2 400 MHz band under Licensed Shared Access (LSA)},
2014.


\bibitem{ETSIFM52}
Faussurier, E., ANFR, Chairman CEPT/WGFM Project Team FM53 
\emph{The potential of Licensed Shared Access for the wireless broadband growth - Regulatory aspects of LSA}, 
\url{http://www.fub.it/sites/default/files/attachments/2014/02/Faussurier.pdf}.

\bibitem{Rep56}
CEPT Report 56,
\emph{Technological and regulatory options facilitating sharing
between Wireless broadband applications (WBB) and the
relevant incumbent services/applications in the 2.3 GHz
band},
2015.

\bibitem{CRANoverview}
Checko, A.; Christiansen, H.L.; Yan, Y.; Scolari, L.; Kardaras, G.; Berger, M.S.; Dittmann, L.,
\emph{Cloud RAN for Mobile Networks—A Technology Overview},
IEEE Communications Surveys \& Tutorials,
2015.

\bibitem{NWOB}
Doyle, L.; Kibilda, J.; Forde, T.K.; Dasilva, L.,
\emph{Spectrum Without Bounds, Networks Without Borders},
Proceedings of the IEEE , vol.102,
2014.

\end{thebibliography}

\end{document}